\documentclass[useAMS,usenatbib]{mn2e}
\usepackage{times}
\usepackage{natbib}
\usepackage{aas_macros}
\usepackage{amsmath}
\usepackage[pdftex]{graphicx}

\title[]{Rise and Fall of Radio Halos in Simulated Merging Galaxy Clusters}
\author[J. Donnert, K.Dolag, G.Brunetti, R.Cassano]{
J. Donnert$^{1,2}$\thanks{donnert@ira.inaf.it}, K.Dolag$^{2,3}$, G.Brunetti$^{1}$, R.Cassano$^{1}$\\
$1$ INAF Istituto di Radioastronomia, via P. Gobetti 101, I-40129 Bologna, Italy\\
$2$ Max Planck Institute for Astrophysics, P.O. Box 1317, D--85741 Garching, Germany\\
$3$ Universit\"atssternwarte M\"unchen, Scheinerstr. 1, D-81679 M\"unchen, Germany\\
}

\begin{document}

\date{Accepted ???. Received ???; in original form ???}

\pagerange{\pageref{firstpage}--\pageref{lastpage}} \pubyear{2012}

\maketitle

\label{firstpage}

\begin{abstract}
We present the first high resolution MHD simulation of cosmic-ray electron reacceleration by turbulence in cluster mergers. We use an idealised model for cluster mergers, combined with a numerical model for the injection, cooling and reacceleration of cosmic-ray electrons, to investigate the evolution of cluster scale radio emission in these objects. In line with theoretical expectations, we for the first time, show in a simulation that reacceleration of CRe has the potential to reproduce key observables of radio halos. In particular, we show that clusters evolve being radio loud or radio quiet, depending on their evolutionary stage during the merger. We thus recover the observed transient nature of radio halos. In the simulation the diffuse emission traces the complex interplay between spatial distribution of turbulence injected by the halo infall and the spatial distribution of the seed electrons to reaccelerate. During the formation and evolution of the halo the synchrotron emission spectra show the observed variety: { from power-laws with spectral index of 1 to 1.3 to curved and ultra-steep spectra with index $> 1.5$}. 
\end{abstract}


\section{Introduction}\label{intro}

Galaxy clusters form in the knots of the cosmic web through the infall and merging of smaller structures. The gravitational potential of these objects is dominated by dark matter (DM). The intra-cluster-medium (ICM) forms inside this potential well as a complex mixture of thermalised plasma and non-thermal components - magnetic fields and cosmic-rays. Non-thermal emission is observed in the radio band from clusters \citep[i.e.][]{2012A&ARv..20...54F}, most spectacular in the form of giant radio halos. 
The origin of the underlying synchrotron bright cosmic-ray electrons (CRe) is still unclear, although the two commonly discussed mechanisms are reacceleration by merger-driven turbulence \citep{2001MNRAS.320..365B,2001ApJ...557..560P,2002ApJ...577..658O,2003ApJ...584..190F,2005MNRAS.357.1313C} and the in-situ production of CRe by proton-proton collisions \citep{1980ApJ...239L..93D,1999APh....12..169B,2000A&A...362..151D,2004A&A...413...17P} or a combination of the two mechanisms \citep{2005MNRAS.363.1173B,2011MNRAS.412..817B}. To date, pure secondary models appear disfavoured by radio spectra of some halos \citep{1987A&A...182...21S,2003A&A...397...53T,2003ApJ...588..155R,2008Natur.455..944B,2010A&A...517A..43M,2010MNRAS.401...47D}. {  In addition $\gamma$-rays are unavoidably produced by the same decay chain as the CRe in these models, but to date, clusters remain not observed in this regime \citep{2009A&A...502..437A,2009ApJ...704..240K,2010ApJ...717L..71A}. This leaves hadronic models assailable, assuming cluster magnetic fields derived from the analysis of Faraday rotation measures  \citep{2011ApJ...728...53J}. Several works attempt to circumvent these difficulties by using a wider range of model parameters \citep{2012ApJ...757..123A,2012arXiv1207.6410Z}. However a more detailed analysis that combines self-consistently {\it state-of-the-art} radio observations of the Coma radio halo with the corresponding $\gamma$-ray (FERMI) upper limits  challenges a pure hadronic origin of CRe in this prototypical radio halo \citep{2012MNRAS.426..956B}. } \par
 
Simulations of galaxy clusters have been used for a decade now to study the evolution of turbulence on the largest scales \citep{2001A&A...378..777D}. This has been done in the cosmological \citep{2005JCAP...01..009D,2007ApJ...655...98N,2008Sci...320..909R,2009A&A...504...33V,2011MNRAS.414.2297I} and the idealised context \citep{1999ApJ...518..594R,2005ApJ...629..791T}. Here specifically idealised simulations of cluster mergers usually neglect substructures of the two systems and the background expansion of the universe, but merge two spherically symmetric halos \citep{2001ApJ...561..621R,2008ApJ...687..951T,2011ApJ...743...16Z}. These simulations allow for full control of the system parameters and present a viable step towards fully cosmological approaches, which offer less control but more realism. \par 
Simulations of CRs in galaxy clusters have so far been limited to studies involing CR protons and the injection of CRe in shocks \citep{2001ApJ...562..233M,2008MNRAS.385.1211P,2011A&A...529A..17V,2012MNRAS.421.3375V}. These simulations did not include a treatment of turbulent reacceleration  and were in fact unable to explain the main observed properties of giant radio halos simultaneously, mostly because of the underlying CR model \citep{2010MNRAS.401...47D,2010MNRAS.407.1565D,2011MNRAS.412....2B}. \par

In this paper, we for the first time attempt to include the complex physics of CRe reacceleration by merger driven turbulence into simulations. This must be considered a first step in this direction, as we model only CRe and simulate only an idealised binary collision. Still we are able to follow the evolution of  turbulence, magnetic fields, CRe and the subsequent radio synchrotron emission during mergers. We use a model for idealised cluster collisions based on an analytic solution of the hydrostatic equation in a Hernquist potential (Donnert in prep.). A low viscosity, high order description for SPH allows us to simulate the flow with high Reynolds numbers to follow the rise and decay of turbulence in the cluster. We then compute the evolution of CRe in post-processing to the simulation considering all relevant losses as well as for the first time stochastic reacceleration of CRe. Synthetic observations are then extracted using a numerical solver for the synchrotron integral. \par
This paper is organised as follows: in section \ref{crmodel} we summarize the CRe model assumed in our simulation. Section \ref{method} describes the numerical method used to integrate the Fokker-Planck equation, treat turbulence in SPH and build the initial conditions. We present our results in section \ref{results} and discuss these in \ref{disc}.

\section{Cosmic-ray Model}\label{crmodel}

In this paper, we consider only CRe and assume that they are re-accelerated by large scale MHD turbulence following \citet{2007MNRAS.378..245B}. There is agreement on the fact that turbulence in the ICM cannot efficiently accelerate thermal particles \citep{2008ApJ...682..175P}. However that mechanism is efficient enough to re-energize relativistic seed electrons \citep{2001ApJ...557..560P,2001MNRAS.320..365B,2002ApJ...577..658O}. Consequently, following previous works in the field, we assume a population of seed non-thermal electrons supplied by $Q_{\mathrm{e}}(p,t)$. As this is an explorative study, we do not elaborate a physical model of the injection process of CRe. Possible { injection processes} are diffusive shock acceleration, hadronic processes, reconnection or starbursting galaxies and AGN \citep{2007MNRAS.375.1471B,2011MmSAI..82..636L}. { It has been realised early on that spatial diffusion of CRe is a slow process \citep[see e.g.][ for a review]{2007IJMPA..22..681B}. We therefore neglect it in this work.}

Considering an isotropic spectral energy distribution of CRe, $n(p)$, the equation governing the evolution of CRe is then:
\begin{align}
    \frac{\partial n(p,t)}{\partial t} &= \frac{\partial}{\partial p}\left[ n(p,t)\left( \left.\frac{\mathrm{d}p}{\mathrm{d}t}\right|_{\mathrm{rad}} + \left.\frac{\mathrm{d}p}{\mathrm{d}t}\right|_{\mathrm{i}} -\frac{2}{p} D_{\mathrm{pp}}(p) \right) \right] \nonumber\\
    &+ \frac{\partial}{\partial p}\left[ D_{\mathrm{pp}}(p) \frac{\partial n(p,t)}{\partial p} \right] + Q_{\mathrm{e}}(p,t),\label{eqn.fkp}
\end{align}
where $\frac{\mathrm{d}p}{\mathrm{d}t}$ are the standard radiative and ionisation losses \citep{2005MNRAS.357.1313C}: 
\begin{align}
    \left.\frac{\mathrm{d}p}{\mathrm{d}t}\right|_{\mathrm{rad}} &= -4.8\times10^{-4} p^2 \left[ \left( \frac{B_{\mu G}}{3.2} \right)^2 + 1\right],\label{eq.radlosses}\\
    \left.\frac{\mathrm{d}p}{\mathrm{d}t}\right|_{\mathrm{i}} &= -3.3\times10^{-29}n_{\mathrm{th}}\left[ 1 + \mathrm{ln}\left(\frac{\gamma}{n_{\mathrm{th}}}\right)/ 75 \right],\label{eq.ilosses}
\end{align}
where $\gamma = p_e/m_{e}c$ is the Lorentz factor and $n_{\mathrm{th}}$ the number density of thermal gas. {  Specifically, we model the injection coefficient as:
\begin{align}
Q_{\mathrm{e}}(\gamma,t) = K_e \varepsilon_\mathrm{th}^2\gamma^{-2} F\left(\gamma\right) \label{eq.inj},
\end{align} 
where $\varepsilon_\mathrm{th}$ is the thermal energy density of the plasma and $F\left(\gamma\right)$ is a cut-off function that limits the injection to energies of $\gamma \in [50, 10^5]$ . This leaves $K_e$ the only free parameter in equation \ref{eqn.fkp}.  Eq. \ref{eq.inj} would eventually mimic continuous injection of secondary particles by inelastic collisions between CRp and thermal protons, with CRp following the spatial distribution of the thermal matter. For $D_\mathrm{pp}=0$ in equation \ref{eqn.fkp}, this results in the asymptotic solution: }
\begin{align}
    n(p) &=  \frac{1}{\left.\frac{\mathrm{d}p}{\mathrm{d}t}\right|_{\mathrm{rad}} + \left.\frac{\mathrm{d}p}{\mathrm{d}t}\right|_{\mathrm{i}}}  \int\limits_\mathrm{p} Q(p) \,\mathrm{d}p \label{eq.steadystate}
\end{align}
{  that is $n(p) \propto p^{-3}$ at high energies, where losses are dominated by synchrotron radiation and IC scattering (eq.\ref{eq.radlosses}).}
To be conservative, we use only one mechanism to reaccelerate particles in the turbulent ICM, that is the interaction with fast modes due to the Transit-Time-Damping resonance \citep{1998ApJ...492..352S}. In this limit, the reacceleration coefficient  considering the relevant collisionless damping of compressive turbulence in the ICM \citep{2007MNRAS.378..245B}  can be simplified to :
\begin{align}
    \frac{D_{\mathrm{pp}}}{p^2} &  \approx 10^{-7} \frac{c_{\mathrm{s}}}{l} \left( \frac{v_{\mathrm{turb}}}{c_{\mathrm{s}}} \right)^4 \eta^2   , \label{eq.dpp}
\end{align}
where we assume a fraction $\eta = 0.45$ of turbulent energy in fast modes, $c_\mathrm{s}$ is the sound speed, and $v_\mathrm{turb}$ is the velocity of turbulent\footnote{for a Kolmogorov spectrum: $v_\mathrm{turb} \propto l^{1/3}$} eddies at the scale $l$.

\section{Numerical Method}\label{method}
We use the MHDSPH code {\small GADGET3} \citep{2005MNRAS.364.1105S,2009MNRAS.398.1678D} to follow collisionless and gas dynamics of DM and the ICM plasma in the MHD approximation.  \par

\subsection{Initial Conditions}\label{ics}

\begin{table}
    \centering
    \begin{tabular}{c c l l}
       Value & Unit & Cluster 0 & Cluster 1 \\\hline
       $\beta$ & - & 2/3 & 2/3\\
       $a_\mathrm{Hernq}$ & kpc & 811  & 473 \\
       $r_{200}$ & kpc & 2192 & 1380 \\
       $r_\mathrm{core}$  & kpc & 237 & 130 \\
       $\mathrm{M}_{200}$ & $10^{15} \mathrm{M}_\odot $ & 1.33 & 0.17 \\
       $\rho_\mathrm{0,gas}$&  $10^{-26}\mathrm{g}/ \mathrm{cm}^3$ & 0.9 & 1.1 \\
       $T(r_\mathrm{core})$ & $10^7 \mathrm{K}$ & 9.2  &  2.8 
    \end{tabular}
    \caption{Model parameters of the initial conditions based on Donnert in prep. }\label{tab.ic}
\end{table}

We model the collision of two dark matter halos using Hernquist profiles, which can be identified with a NFW profile in the core \citep{2007MNRAS.380..911S}. The system has a total mass of $1.5\times10^{15}\,\mathrm{M}_\odot$, with a mass ratio of 1:8. We chose a comparatively low mass-ratio in this simulation to reduce the influence of the shock phase to the radio emission. We leave a more general investigation to future work. The density profiles are sampled by DM and SPH particles, $10^7$ each. From the mass $M_{200}$ in $r_{200}$ we obtain the Hernquist scale length using the observed relation of the concentration parameter $c_\mathrm{NFW} \approx 3$ \citep{2008MNRAS.390L..64D}. For the baryonic matter, we adopt the widely used $\beta$-model, with $\beta=2/3$  and $r_\mathrm{core} = 0.3 r_{200}/c_\mathrm{NFW}$ { with a baryon fraction of 0.17} \citep{2001ApJ...561..621R,2008MNRAS.389..967M}. For these parameters, the hydrostatic equation can then be solved analytically (Donnert in prep.). {  Similar to the observations of non cool-core clusters, the temperature profile is then almost constant in the central part and mildly declines towards the outer parts, with a mean temperature within the core region as listed in table \ref{tab.ic}, alongside other details of the system.} \par
The two clusters are set-up separately, without magnetic field, and relaxed for several Gyrs. The relaxation process is done with high viscosity and thermal conduction \citep{2004MNRAS.351..423J}, to quickly equalise density fluctuations induced by the Poisson sampling of the gas density distribution. \par
In a second step, the clusters are joined in a periodic box of 10 Mpc. { This allows a consistent treatment of the magnetic field in k-space. Furthermore, periodicity prevents low timesteps on escaping SPH particles.} The two halos are put on a zero energy orbit at a distance so that the density of both profiles match with an impact parameter of 300 kpc. Overlapping particles are resampled to the volume outside of $r_{200}$. { The magnetic field is then initialised as a constant divergence free vector-field in k-space, with a Kolmogorov power-spectrum\footnote{as measured by e.g. \citet{2011A&A...529A..13K} in Hydra A} down to 150 kpc. This minimises the oversampling of the SPH kernel in low density regions, which is known to cause numerical divergence. The vector-field is then transformed to real-space and sampled to the particles with the TSC kernel. The resulting field is rescaled to a maximum value of $3\,\mu \mathrm{G}$ in the cluster centre and attenuated radially as the square-root of the gas density\footnote{as derived from Faraday rotation measurements by \citet{2010A&A...513A..30B}}. Residual magnetic divergence is then treated by the divergence cleaning in the code at runtime (see section \ref{turb}).}

\subsection{Treatment of Turbulence}\label{turb}

\begin{figure}
\centering
\includegraphics[width=0.45\textwidth]{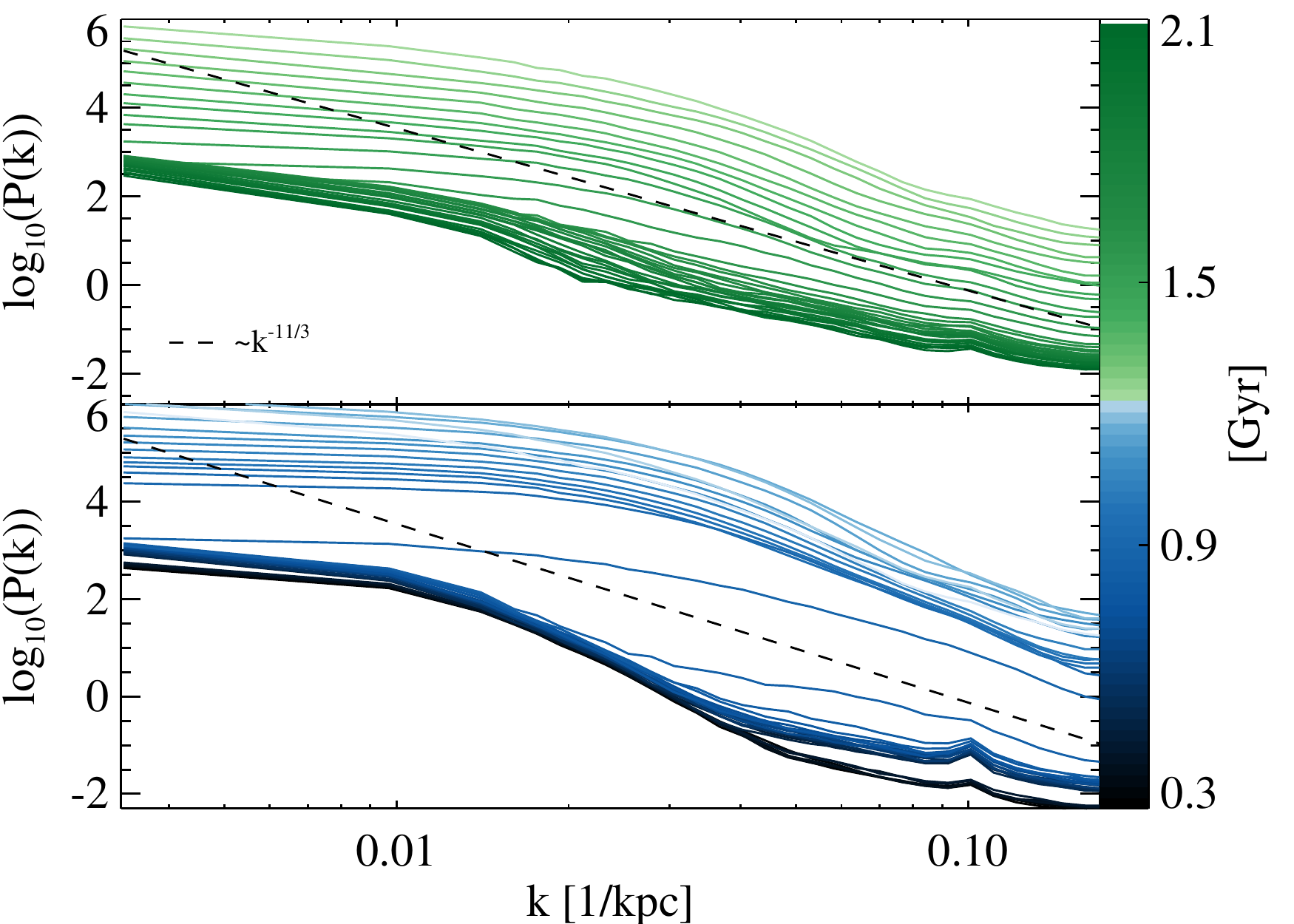}
\caption{Velocity power-spectra of the inner most region of the system at different times. Top: after maximum halo brightness, bottom: before maximum halo brightness. Time evolution is color coded. The dashed lines follow the Kolomogorov scaling of $k^{-11/3}$.} \label{img.Pk}
\end{figure}

To follow the turbulent reacceleration of the CRe, we need to trim the SPH formalism to follow the generation of turbulent motions and their propagation, possibly over a broad range of scales within the simulation. Furthermore, we have to estimate the  turbulent energy at every location within the cluster during the merger event. Therefore, we use a largely improved SPH scheme, where: 
\begin{enumerate}

\item Instabilities can be resolved by using a time-dependent formulation of thermal conduction \citep{2008JCoPh.22710040P}, which allows a mixing similar to grid codes. This additionally eliminates discontinuities in our initial conditions.  

\item We employ a time-dependent viscosity scheme based on the  local signal velocity \citep[e.g.][]{2005MNRAS.364..753D}, which significantly reduces the numerical viscosity outside of shocks.

\item We use a higher order kernel based on the bias-corrected fourth  order Wendland kernel \citep{2012arXiv1204.2471D} with 200 neighbors. This corresponds to the same effective kernel FWHM (smoothness) as the standard, cubic kernel with 64 neighbours, but increases the compact support of the kernel considerably \citep{2012arXiv1204.2471D}. Then dissipation takes place almost entirely inside this kernel scale.
\end{enumerate}

This way the SPH algorithm formally models a flow, with a Reynolds number of more than 1000 away from shocks, and sufficiently follows turbulent motions up to the kernel scale. In such an implementation, SPH does form a cascade down to small scales \citep{2012arXiv1206.5006H}, where turbulent motions are eventually thermalised and damped by the viscosity scheme.  This approach is therefore conservative in terms of turbulent energy. \par

In figure \ref{img.Pk}, we show the filtered velocity power-spectrum of a region of $1 \, \mathrm{Mpc}^3$ around the center of mass of the system before, during and after the merger. Here we binned the SPH particle velocity to a grid. Then we subtracted the mean velocity in a region of 100 kpc around each cell to filter out the bulk motion of the system.\par
{ The intial power-spectrum shows fluctuations of the order of 10km/s, due to the underlying numerical algorithm.} The amplification of the velocity power during the passage is indicative of a drastic increase in turbulent motions in that phase. Furthermore, the subsequent decay of that velocity power then demonstrates the turbulent origin of these motions and their dissipation on small scales. Thus, we use a high order kernel which employs the same effective size, but a much larger region and number of neighbors than standard approaches. This allows us to estimate the local turbulent energy, similar to sub-grid turbulence models for modelling unresolved turbulence in galaxy clusters  \citep{2011MNRAS.414.2297I}. Here we use the RMS velocity fluctuation filtered on the kernel scales. { This implies $l = 2 h_{\mathrm{sml}}$ in equation \ref{eq.dpp}, where $h_{\mathrm{sml}}$ is the compact support of the SPH kernel.} Note that compared to grid codes, this method typically underestimates the turbulent energy on the kernel scale.\par

{  We model a turbulent magnetic resistivity, following \citet{2011MNRAS.418.2234B}. This also acts as a cleaning scheme for magnetic divergence in turbulent flows and treats residual divergence in the outer regions as well as the initial conditions \citep{2012JCoPh.231..759P}. }

\subsection{Cosmic-Ray Electrons}\label{fkp}

\begin{figure}
\centering
\includegraphics[width=0.45\textwidth]{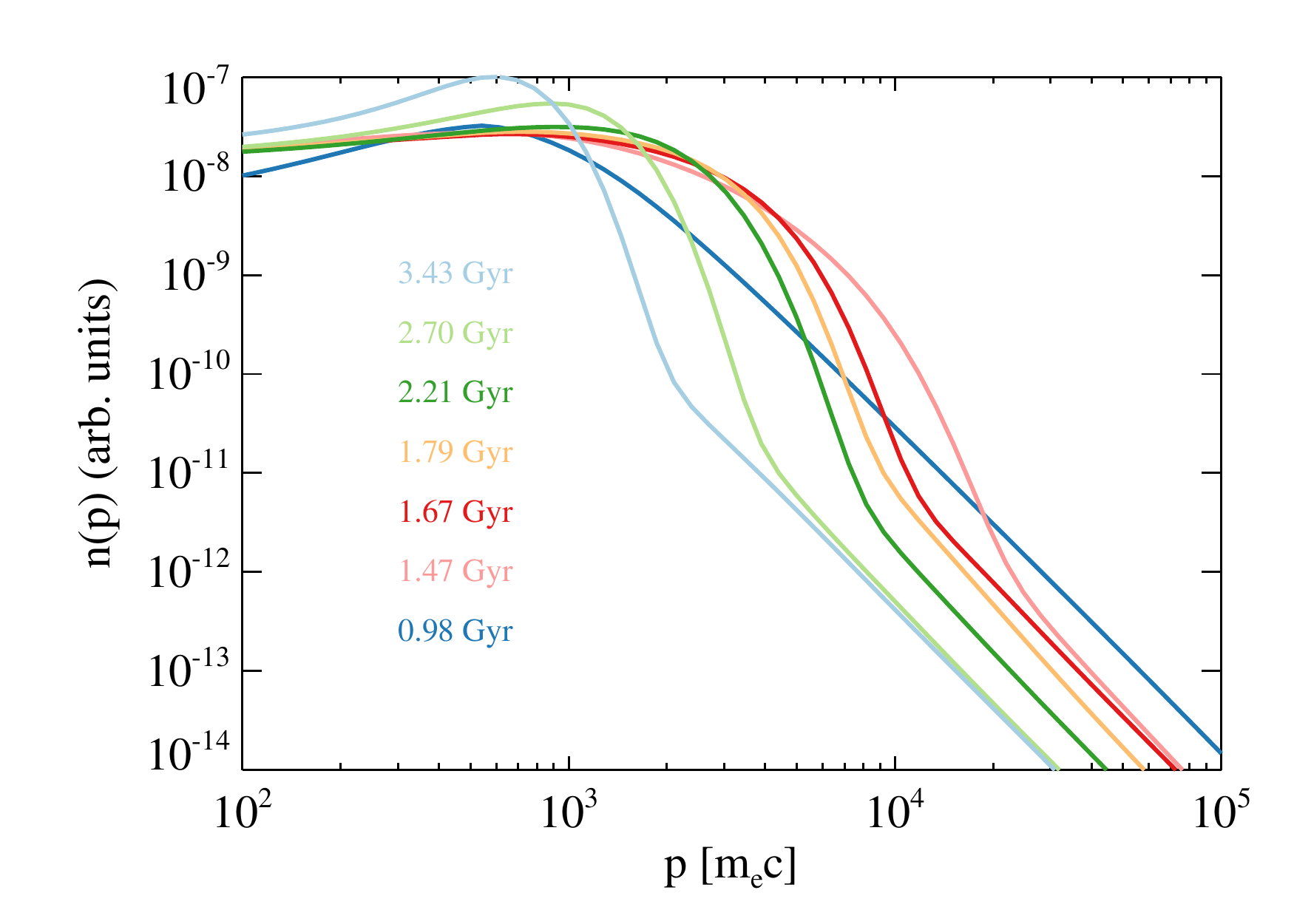}
\caption{Spectrum of an exemplary particle (ID 400383) at different times during the merger. Colours correspond to figure \ref{img.correlation} and \ref{img.spectrum}.} \label{img.CreSpec}
\end{figure}

We solve equation \ref{eqn.fkp} for every SPH particle, in post-processing to the simulation. We employ the adaptive upwind scheme developed by \citet{1970CompPhys.ChangCooper}. Per particle, we logarithmically sample the isotropic number density spectrum of CRe 100 times in the range of $p/m_e c \in [0.1, 10^5]$, with open boundary conditions realised, as in \citet{1986ApJ...308..929B}. The fluid quantities are interpolated linearly between 290 snapshots every 25 Myr. We employ a universal timestep of 0.1 Myr. { For example, we show the spectrum of one particle in figure \ref{img.CreSpec}. It shows the effect of turbulent acceleration on the spectrum of the CRe. Initially (dark blue), the system shows the asymptotic solution (eq. \ref{eq.steadystate}), then (light red, red) turbulent acceleration modifies the spectrum up to $\gamma \ge 3\times 10^4$. At the highest energies, the spectrum is shaped by the constant injection of particles (eq. \ref{eq.inj}). The norm of this injection varies with time as the underlying thermal conditions evolve. At late stages (green, light green, light blue), turbulent acceleration is less efficient, leading to the accumulation of particles at synchrotron dark energies around $\gamma < 10^3$.} \par
To produce synthetic observations from our simulations, we use our  MPI-parallel projection code {\small SMAC2}, and numerically solve the synchrotron integral for every particle. 

\section{Results}\label{results}
\begin{figure*}
\includegraphics[type=jpg,read=.jpg,ext=.jpg,width=0.8\textwidth]{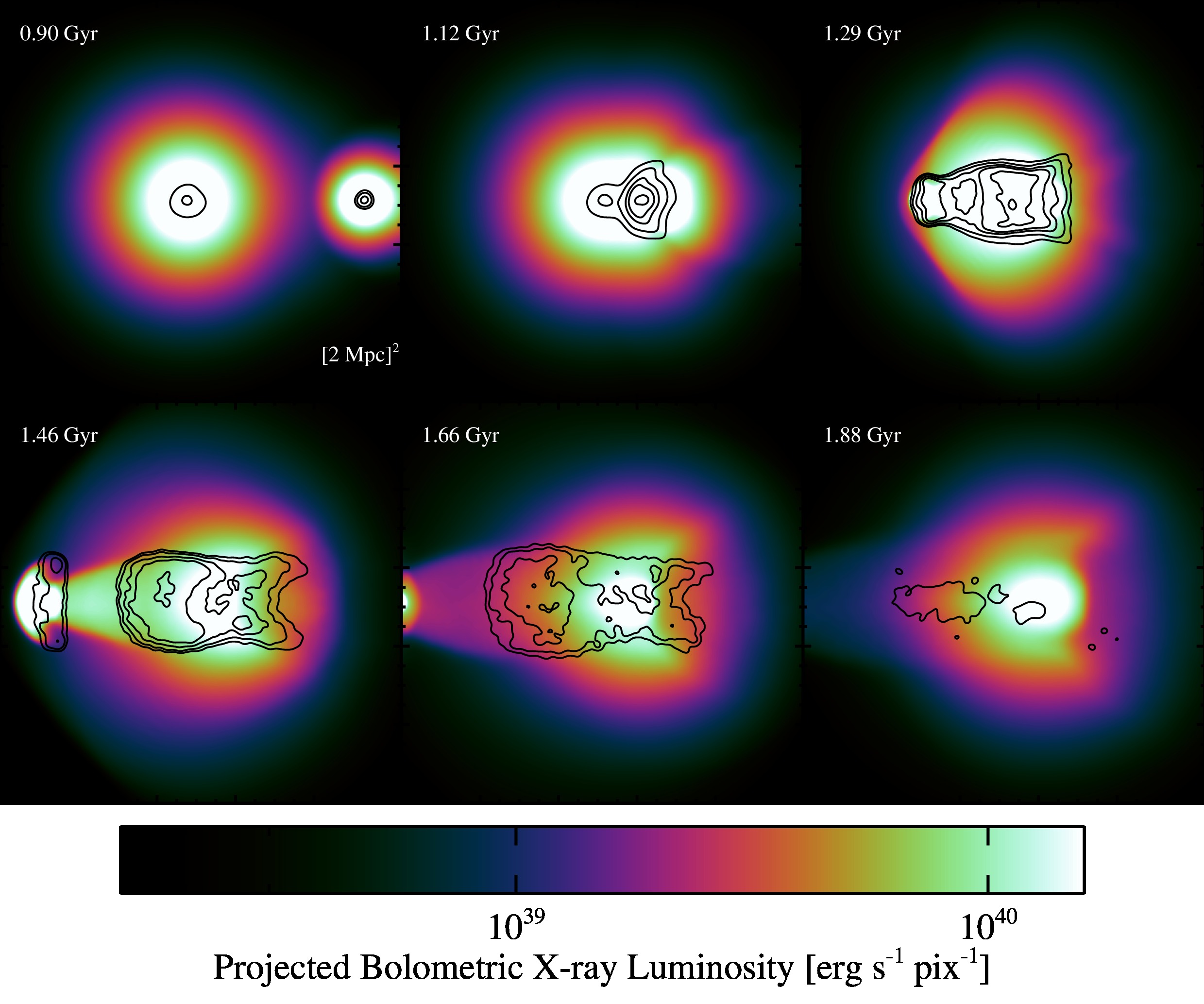}
\caption{Projections of { bolometric} X-ray luminosity of the system rotated by 45. Overlayed radio contours { (1.4GHz)} are at $0.050, 0.22, 7.3, 1.45 \times 10^{30}$ $\mathrm{erg}\, \mathrm{s}^{-1}$.}\label{img.contours}
\end{figure*}
The simulation is evolved for roughly 4 Gyr, and CRe spectra and synthetic observations are obtained. The first core-passage occurs at $\approx 1.18$ Gyrs. In figure \ref{img.contours}, we show X-ray emission of the system at different times. \par 

Despite the simulation being highly idealized, which is reflected in morphological details (high symmetry and regular shapes in the X-rays), the system can be classified into relaxed (before) and disturbed (after) state. We find that Mpc-scale synchrotron radio emission is generated as a result of particle acceleration by merger turbulence, in connection with the collision between clusters.  The radio - X-ray evolution of the system can be summarized in 3 phases: infall and shock dominance, reacceleration phase and decay.
\begin{enumerate}
    \item During \emph{infall} (less than 1.25 Gyr after the start of the simulation), X-ray and radio luminosity increase, within a few hundred Myr, by a factor of 2 and 30, respectively. The radio emission is localized at the region crossed by the shock front.
    \item The system enters the \emph{reacceleration} phase shortly after the core passage  ($1.25 < t < 1.7$ Gyr):  The X-ray brightness rapidly declines due to the decrease in density, caused by the disturbance of the core. In contrast, the total radio synchrotron brightness continues to increase, as the DM core drives turbulence in a large fraction of the cluster volume. 
    \item During the \emph{decay} ($t > 1.7$ Gyr)  phase, the radio emission traces the trail of the turbulence driven by the secondary DM core, fades away gradually and becomes offset from the primary core.
\end{enumerate}

\begin{figure}
\includegraphics[width=0.45\textwidth]{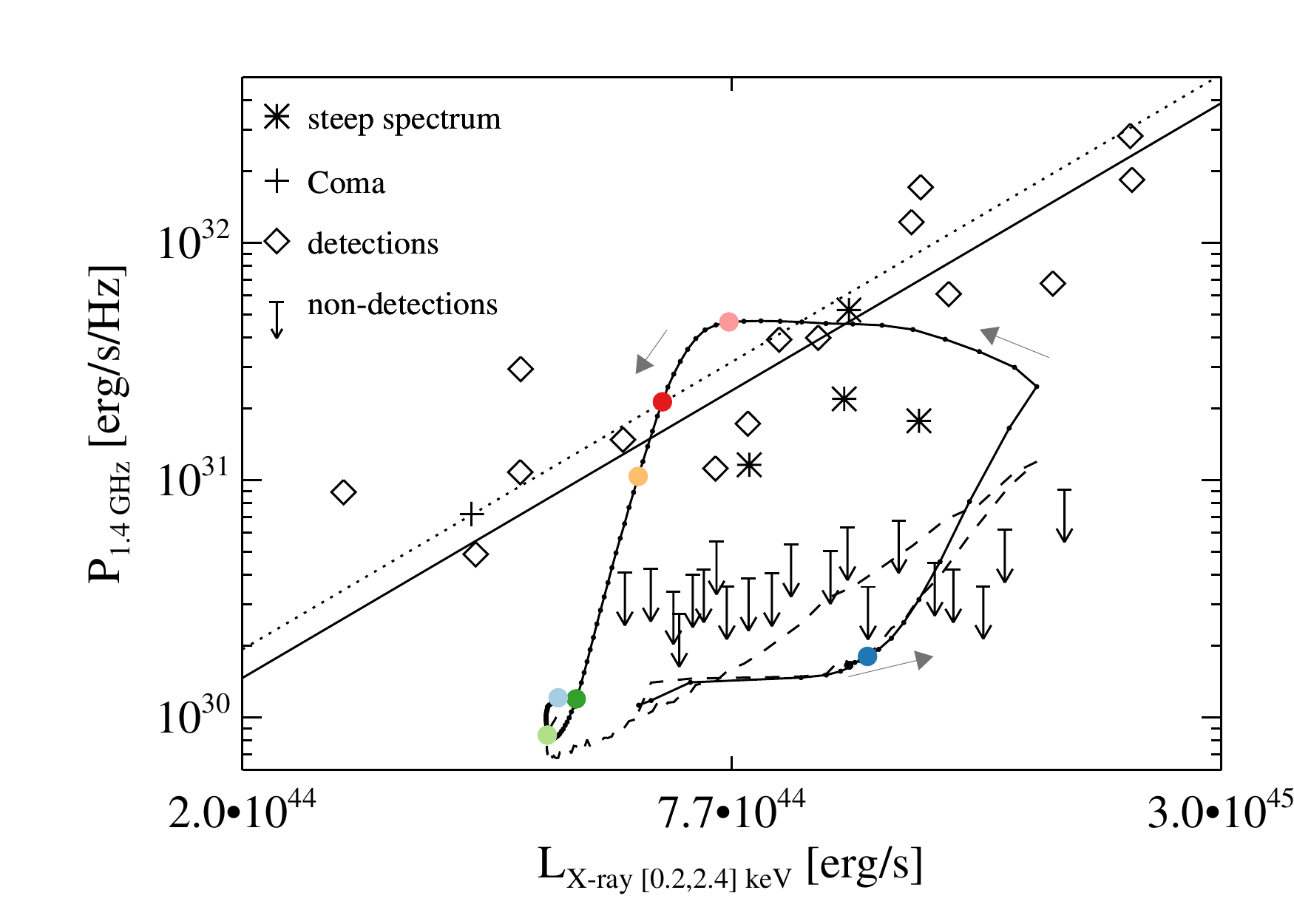}
\caption{Evolution of the system in X-rays and radio synchrotron emission as black line. We mark the time evolution as dots on the lines at an interval of 25 Myr. The colored dots correspond to the times shown in figure \ref{img.spectrum}. The dashed line shows the emission from the injection only. Overplotted are observed radio halos and upper limits \citep{2009A&A...507..661B} as well as the correlation (black line). To guide the eye the correlation is scaled to Coma (black dotted). We mark Coma with a black cross. We also include known upper limits as arrows. Ultra steep spectrum halos A521, A1914, MACSJ1149.5+2223 and A697  are plotted as asterisks \citep{2009ApJ...699.1288D,2011A&A...533A..35V, 2012arXiv1206.6102B}.}\label{img.correlation}
\end{figure}

\begin{figure}
\includegraphics[width=0.45\textwidth]{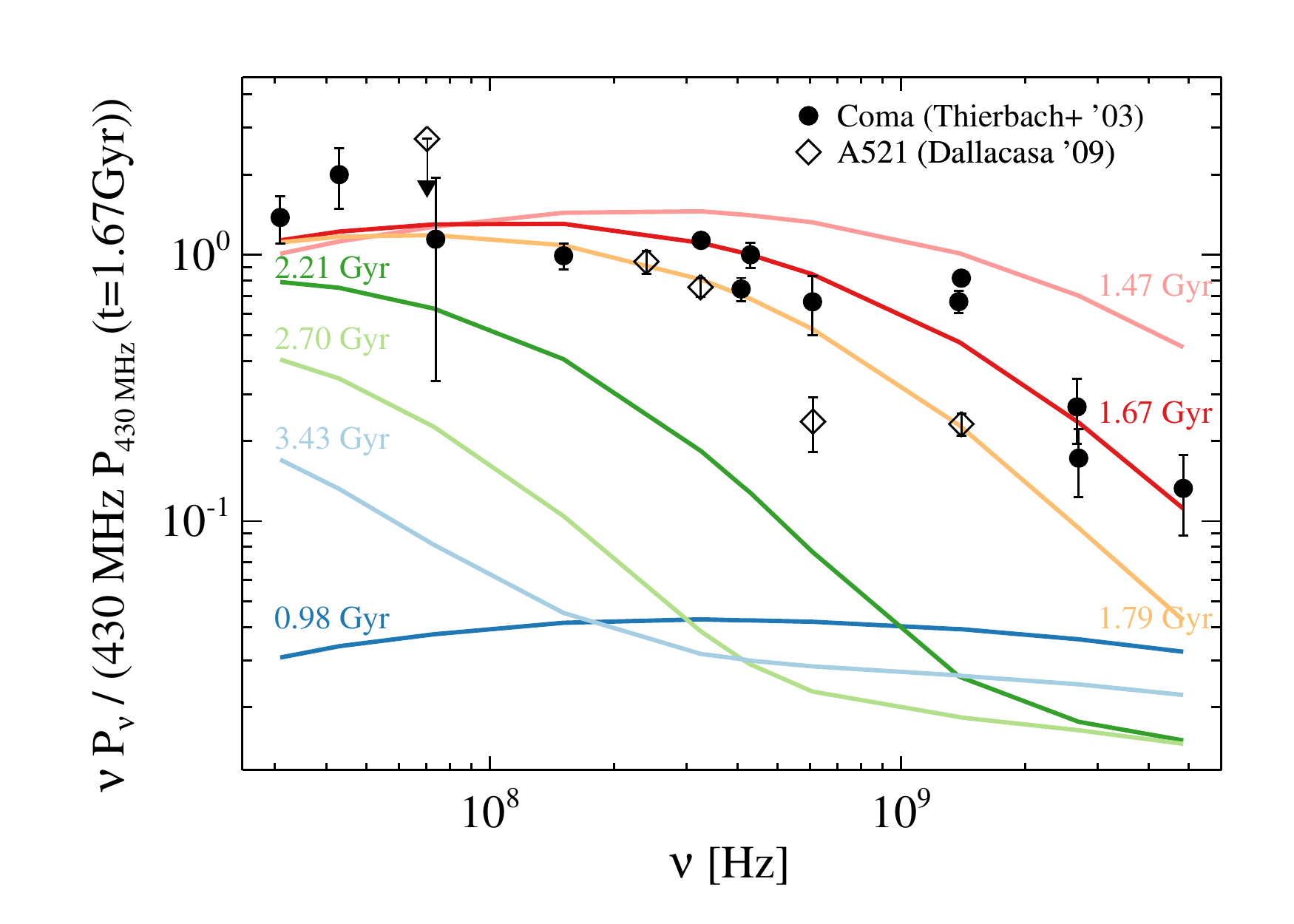}
\caption{Radio synchrotron spectrum of the system at different times. Observed spectrum of the Coma cluster as black diamonds from \citet{2003A&A...397...53T}. Observed spectrum from A521 as asterisks \citep{2009ApJ...699.1288D}.}\label{img.spectrum}
\end{figure}

Present radio surveys suggest that radio halos are transient phenomena connected to cluster mergers \citep{2010ApJ...721L..82C}. In figure \ref{img.correlation}, we plot the evolution of the system in the Lx - P14 plane as a black line. { This is compared with observed radio halos (astrisk - ultra-steep spectrum) and present upper limits.} We mark time intervals of 25 Myr as dots on the line, and the times shown in figure \ref{img.spectrum} as large dots with corresponding colour. The emission from the injection only, i.e. without considering turbulent reacceleration, is overplotted as dashed line. { To determine the CRe normalisationi, a time was chosen, so that the simulated spectrum fits the observed Coma spectrum. Then the normalisation was set to $K_\mathrm{e} = 2 \times 10^{-4}$, so that the halo brightness roughly fits the Coma luminosity extrapolated along the correlation (dotted line in figure \ref{img.correlation}). This corresponds to a total energy injected in the form CRe during our simulation of:}
\begin{align}
    \frac{\varepsilon_\mathrm{CRe,inj}}{\varepsilon_\mathrm{th}} & \approx 10^{-3} \left(\frac{\tau_\mathrm{inj}}{2\,\mathrm{Gyr}}\right) \left( \frac{<\varepsilon_\mathrm{th}>}{2\times10^{-11}\,\frac{\mathrm{erg}}{\mathrm{cm}^3}}\right),
\end{align}
where $<\varepsilon_\mathrm{th}>$ is the typical thermal energy density and $\tau_\mathrm{inj}$ is the injection time.

Before the first passage, the radio luminosity of the system is more than a factor of 10 below the correlation. This corresponds to the "off-state" of galaxy clusters without an observed radio halo, i.e. the upper limits from observations. During the shock phase of the first passage, X-ray and radio luminosity of the system increase and the system approaches the zone of radio halos. In the reacceleration phase, the system crosses this zone as the X-ray luminosity declines rapidly by a factor of two.  The radio luminosity decreases, within 1 Gyr, to injection values, i.e. the system enters the radio-quiet phase. The system spends $\sim0.7$ Gyrs inside the observed scatter around the correlation and $\sim0.17$ Gyrs to move between correlation and upper limits. 

The spectral properties of the system change drastically during its evolution in the P14-LX plane. In figure \ref{img.spectrum}, we plot the radio spectrum of the system at different times. We overplot the radio spectrum of the COMA halo \citep[see][ and ref. therein]{2003A&A...397...53T} as well as A521 \citep{2009ApJ...699.1288D}.  The simulated spectra of Coma and A521 are scaled to match the observed ones at 430 MHz and 240 MHz, respectively.\par
Before the merger, the system shows only emission with a spectral index of $\approx -1$, consistent with {  the steady-state spectrum due to losses and constant injection (eq. \ref{eq.radlosses}, \ref{eq.ilosses}, \ref{eq.inj})}. At maximum luminosity, the radio spectrum fits the observed COMA spectrum.  A comparison with A521  yields a reasonable fit of the spectrum at 1.8 Gyr after start of the simulation. Over time, the total emission decreases and further steepens at high frequencies.  Different assumptions for the fraction of compressive modes would lead to systematically flatter (high $\eta$) or steeper (low $\eta$) spectra. However, for the first time, we have shown that viable assumptions for this parameter in simulations enable us to naturally explain the variety of observed spectra \citep{2011MmSAI..82..499V} in giant radio halos. \par

\section{Conclusions}\label{disc}

We obtained, for the first time, an MHD simulation of two colliding, idealized galaxy clusters, including the evolution of the magnetic field as well as a combined treatment  of turbulence and CR electrons  reacceleration within the ICM\footnote{We note that a similar approach was adopted for mini-halos and core-sloshing by  \citet{2012arXiv1203.2994Z}}. Reacceleration is treated by a simple conservative model, assuming TTD damping of fast magneto-sonic waves by relativistic particles. The simulated evolution of the ICM and all relevant cooling mechanisms are coupled self-consistently to the CRe. We numerically solve the underlying Fokker-Planck equation for all $10^7$ gas particles of the cluster merger simulation, sampling the CRe distribution function by 100 logarithmically-spaced energy bins, in the  range $\gamma=[0.1,10^5]$. This allows us to obtain synthetic radio observations in the observed frequency range (30MHz - 3GHz) and compare to the distribution and evolution of the X-ray emission of the system. { The inclusion of these techniques comprises a significant step towards a self-consistent treatment of CR physics in large-scale astrophysical simulations.} \par
Despite the idealised set-up of the underlying thermal model, our simulation for the first time reproduces several key observables  of giant radio halos. In particular:
\begin{enumerate}
    \item The transient nature of radio halos and their connection to mergers and merger driven turbulence. 
    \item The variety of observed radio spectra (i.e. flatter, curved and ultra-steep spectrum halos) associated with  different states in the merger evolution.
\end{enumerate}
Our results differ from those of previous simplified numerical models, that were based solely on hadronic collision in the ICM and shock injection of CRe and CRp.\par
These results reaffirm previous theoretical arguments that curved (or very steep) spectra of radio halos, their transient nature and connection with mergers are solid predictions of turbulent reacceleration models \citep{2007MNRAS.378..245B,2011MNRAS.412..817B}. Present results must be considered to be only first steps in this direction, as a more self-consistent model requires the inclusion of CRp, the generation of secondary particles and the interplay with turbulence in a cosmological setup. \par

\section{Acknowledgements}\label{ack}
The authors thank the anonymous referee for useful comments.  The computations where performed at the ``Rechenzentrum der Max-Planck-Gesellschaft'', with resources assigned to the ``Max-Planck-Institut f\"ur Astrophysik''.\par
JD thanks R. Kale, F. Vazza, A. Beck and A. Lazarian for enlightening discussions. JD, GB, RC acknowledge partial support by PRIN-INAF2009 and ASI-INAF I/088/06/0. JD acknowledges support from the EU FP7 Marie Curie programme 'People'.

\bibliographystyle{mn2e} \bibliography{master}

\label{lastpage}
\end{document}